\documentclass[11pt]{article}
\usepackage[a4paper, margin=2cm]{geometry}
\usepackage{mathptmx}
\usepackage[T1]{fontenc}
\usepackage[utf8]{inputenc}
\usepackage{color}
\definecolor{document_fontcolor}{rgb}{0, 0, 0}
\color{document_fontcolor}
\usepackage{bm}
\usepackage{amsmath}
\usepackage{amssymb}
\usepackage{graphicx}
\usepackage{tikz}

\begin{document}
\global\long\def\tht{\vartheta}%
\global\long\def\ph{\varphi}%
\global\long\def\balpha{\boldsymbol{\alpha}}%
\global\long\def\btheta{\boldsymbol{\theta}}%
\global\long\def\bJ{\boldsymbol{J}}%
\global\long\def\bGamma{\boldsymbol{\Gamma}}%
\global\long\def\bOmega{\boldsymbol{\Omega}}%
\global\long\def\d{\mathrm{d}}%
\global\long\def\t#1{\text{#1}}%
\global\long\def\m{\text{m}}%
\global\long\def\v#1{\mathbf{#1}}%
\global\long\def\u#1{\underline{#1}}%

\global\long\def\t#1{\mathbf{#1}}%
\global\long\def\bA{\boldsymbol{A}}%
\global\long\def\bB{\boldsymbol{B}}%
\global\long\def\c{\mathrm{c}}%
\global\long\def\difp#1#2{\frac{\partial#1}{\partial#2}}%
\global\long\def\xset{{\bf x}}%
\global\long\def\zset{{\bf z}}%
\global\long\def\qset{{\bf q}}%
\global\long\def\pset{{\bf p}}%
\global\long\def\wset{{\bf w}}%
\global\long\def\ei{{\bf \mathrm{ei}}}%
\global\long\def\ie{{\bf \mathrm{ie}}}%
\global\long\def\pt{\partial}%
 
\global\long\def\no{\nonumber}%
 
\global\long\def\del{\delta}%
 
\global\long\def\tg{\tau_{g}}%
 
\global\long\def\tba{\bar{\tau}}%
\global\long\def\s{\mathrm{s}}%
\global\long\def\a{\mathrm{a}}%
\global\long\def\f{\mathrm{f}}%

\title{Gaussian processes for data fulfilling linear differential equations}
\author{Christopher G. Albert\\
Max-Planck-Institut für Plasmaphysik, \\
Boltzmannstr. 2, 85748 Garching, Germany\\
albert@alumni.tugraz.at}
\date{\today}

\maketitle

\abstract{A method to reconstruct fields, source strengths and
physical parameters based on Gaussian process regression is presented 
for the case where data are known to fulfill a given linear differential 
equation with localized sources. The approach is applicable to a wide 
range of data from physical measurements and numerical simulations. It is 
based on the well-known invariance of the Gaussian under linear 
operators, in particular differentiation. Instead of using a generic 
covariance function to represent data from an unknown field, the space 
of possible covariance functions is restricted to allow only Gaussian 
random fields that fulfill the homogeneous differential equation. 
The resulting tailored kernel functions lead to more reliable regression 
compared to using a generic kernel and makes some hyperparameters
directly interpretable. 
For differential equations representing laws of physics such a choice 
limits realizations of random fields to physically possible solutions. 
Source terms are added by superposition and their strength estimated 
in a probabilistic fashion, together with possibly unknown 
hyperparameters with physical meaning in the differential operator.}

\section{Introduction}

The larger context of the present work is the goal to construct reduced
complexity models as emulators or surrogates that retain mathematical
and physical properties of the underlying system. Similar to usual
numerical models, such methods aim to represent infinite systems by
exploiting finite information in some optimal sense. In the spirit
of structure preserving numerics the aim here is to move errors to
the ``right place'', in order to retain laws such as conservation
of mass, energy or momentum.

This article deals with Gaussian process (GP) regression on data with
additional information known in the form of linear, generally partial
differential equations (PDEs). An illustrative application is the
reconstruction of an acoustic sound pressure field and its sources
from discrete microphone measurements. GPs, a special class of random
fields, are used in a probabilistic rather than a stochastic sense:
approximate a fixed but unknown field from possibly noisy local measurements.
Uncertainties in this reconstruction are modeled by a normal distribution.
For the limit of zero measured data a prior has to be chosen whose
realizations take values in the expected order of magnitude. An appropriate
choice of a covariance function or kernel guarantees that all fields
drawn from the GP at any stage fulfill the underlying PDE. This may
require to give up stationarity of the process.

Techniques to fit GPs to data from PDEs has been known for some time,
especially in the field of geostatistics~\cite{Dong1989}. A general
analysis including a number of important properties is given by \cite{van2001kriging}.
In these earlier works GPs are usually referred to as Kriging and
stationary covariance functions / kernels as covariograms. A number
of more recent works from various fields \cite{Graepel2003,Sarkka2011,Raissi2017a}
use the linear operator of the problem to obtain a new kernel function
for the source field by applying it twice to a generic, usually squared
exponential, kernel. In contrast to the present approach, that method
is suited best for source fields that are non-vanishing across the
whole domain. In terms of deterministic numerical methods one could
say that the approach correspond to meshless variants of the finite
element method (FEM). The approach in the present work instead represents
a probabilistic variant of a procedure related to the boundary element
method (BEM), also known as the \emph{method of fundamental solutions
(MFS}) or regularized BEM~\cite{Lackner1976,Golberg1995,Schaback2006}.
As in the BEM, the MFS also builds on fundamental solutions, but allows
to place sources outside the boundary rather than localizing them
on a layer. Thus the MFS avoids singularities in boundary integrals
of the BEM while retaining a similar ratio of numerical effort and
accuracy for smooth solutions. To the author's knowledge the probabilistic
variant of the MFS via GPs has first been introduced by \cite{Mendes2012}
to solve the boundary value problem of the Laplace equation and dubbed
\emph{Bayesian boundary elements estimation method ((BE)$^{2}$M)}.
This work also provides a detailed treatment of kernels for the 2D
Laplace equation. A more extensive and general treatment of the Bayesian
context as well as kernels and their connection to fundamental solutions
is available in~\cite{Cockayne2016} under the term \emph{probabilistic
meshless methods (PMM)}.

While \cite{Mendes2012} is focused on boundary data of a single homogeneous
equation, and~\cite{Cockayne2016} provides a detailed mathematical
foundation, the present work aims to explore the topic further for
application and extend the recent work in~\cite{Albert2019a}. Starting
from general notions some regression techniques are introduced with
emphasis on the role of localized sources. For this purpose Poisson,
Helmholtz and heat equation are considered and several kernels are
derived and tested. To fit a GP to a homogeneous (source-free) PDE,
kernels are built via according fundamental solutions. Possible singularities
(sources) are moved outside the domain of interest. In particular,
boundary conditions on a finite domain can be either supplied or reconstructed
in this fashion. In addition contributions by internal sources are
superimposed, using again fundamental solutions in the free field.
For that part boundary conditions of the actual problem are irrelevant.
The specific approach taken here is most efficient for source-free
regions with possibly few localized sources that are represented by
monopoles or dipoles.

\section{GP regression for data from linear PDEs}

Gaussian processes (GPs) are a useful tool to represent and update
incomplete information on scalar fields\footnote{The more general case of complex valued fields and vector fields is
left open for future investigations in this context.} $u(\v x)$, i.e. a real number $u$ depending on a (multi-dimensional)
independent variable $\v x$. A GP with mean $m(\v x)$ and covariance
function of kernel $k(\v x,\v x^{\prime})$ is denoted as
\begin{equation}
u(\v x)\sim\mathcal{G}(m(\v x),k(\v x,\v x^{\prime})).\label{eq:gp}
\end{equation}
The choice of an appropriate kernel $k(\v x,\v x^{\prime})$ restricts
realizations of ~\eqref{eq:gp} to respect regularity properties
of $u(\v x)$ such as continuity or characteristic length scales.
Often regularity of $u$ does not appear by chance, but rather reflects
an underlying law. We are going to exploit such laws in the construction
and application of Gaussian processes describing $u$ for the case
described by linear (partial) differential equations

\begin{equation}
\hat{L}u(\v x)=q(\v x).\label{eq:Lu=00003Dq}
\end{equation}
Here $\hat{L}$ is a linear differential operator, and $q(\v x)$
is an inhomogeneous source term. In physical laws dimensions of $\v x$
usually consist of space and/or time. Physical scalar fields $u$
include e.g. pressure $p$, temperature $T$ or the electrostatic
potential $\phi_{\mathrm{e}}$. Corresponding laws under certain conditions
include Gauss' law of electrostatics for $\phi_{\mathrm{e}}$ with
Laplacian $\hat{L}=\varepsilon\Delta$, frequency-domain acoustics
for $p$ with Helmholtz operator $\hat{L}=\Delta-k_{0}^{\,2}$ or
thermodynamics for $T$ with heat/diffusion operator $\hat{L}=\frac{\partial}{\partial t}-D\Delta$.
These operators contain free parameters, namely permeability $\varepsilon$,
wavenumber $k_{0}$, and diffusivity $D$, respectively. While $\varepsilon$
may be absorbed inside $q$ in a uniform material model of electrostatics,
estimation of parameters $k_{0}$ or $D$ is useful for material characterization.

For the representation of PDE solutions the weight-space view of Gaussian
process regression is useful. There the kernel $k$ is represented
via a tuple $\bm{\phi}(\v x)=(\phi_{1}(\v x),\phi_{2}(\v x),\dots)$
of basis functions $\phi_{i}(\v x)$ that underlie a linear regression
model
\begin{equation}
u(\v x)=\bm{\phi}(\v x)^{T}\v w=\sum_{i}\phi_{i}(\v x)w_{i}.\label{eq:ux}
\end{equation}
Bayesian inference starting from a Gaussian prior with covariance
matrix $\Sigma_{\mathrm{p}}$ for weights $\v w$ yields a Mercer
kernel
\begin{equation}
k(\v x,\v x^{\prime})\equiv\bm{\phi}^{T}(\v x)\Sigma_{\mathrm{p}}\bm{\phi}(\v x^{\prime})=\sum_{i,j}\phi_{i}(\v x)\Sigma_{\mathrm{p}}^{ij}\phi_{j}(\v x^{\prime}).\label{eq:kbasis}
\end{equation}
The existence of such a representation is guaranteed by Mercer's theorem
in the context of reproducing kernel Hilbert spaces (RKHS)~\cite{Schaback2006}.
More generally one can also define kernels on an uncountably infinite
number of basis functions in analogy to~\eqref{eq:ux} via
\begin{equation}
f(\v x)=\hat{\phi}[w(\bm{\zeta})]=\left\langle \phi(\v x,\bm{\zeta}),w(\bm{\zeta})\right\rangle =\int\phi(\v x,\bm{\zeta})\,w(\bm{\zeta})\,\d\bm{\zeta},
\end{equation}
where $\hat{\phi}$ is a linear operator acting on elements $w(\bm{\zeta})$
of an infinite-dimensional weight space parametrized by an auxiliary
index variable $\bm{\zeta}$, that may be multi-dimensional. We represent
$\hat{\phi}$ via an inner product $\left\langle \phi(\v x,\bm{\zeta}),w(\bm{\zeta})\right\rangle $
in the respective function space given by an integral over $\bm{\zeta}$.
The infinite-dimensional analogue to the prior covariance matrix is
a prior covariance operator $\hat{\Sigma}_{\mathrm{p}}$ that defines
the kernel as a bilinear form
\begin{equation}
k(\v x,\v x^{\prime})\equiv\left\langle \phi(\v x,\bm{\zeta}),\hat{\Sigma}_{\mathrm{p}}\phi(\v x^{\prime},\bm{\zeta}^{\prime})\right\rangle \equiv\int\phi(\v x,\bm{\zeta})\Sigma_{\mathrm{p}}(\bm{\zeta},\bm{\zeta}^{\prime})\phi(\v x^{\prime},\bm{\zeta}^{\prime})\,\d\bm{\zeta}\,\d\bm{\zeta}^{\prime}.\label{eq:k12}
\end{equation}
Kernels of the form \eqref{eq:k12} are known as convolution kernels.
Such a kernel is at least positive semidefinite, and positive definiteness
follows in the case of linearly independent basis functions $\phi(\v x,\bm{\zeta})$~\cite{Schaback2006}.

\subsection{Construction of kernels for PDEs}

For treatment of PDEs possible choices of index variables in \eqref{eq:kbasis}
or \eqref{eq:k12} include separation constants of analytical solutions,
or the frequency variable of an integral transform. In accordance
with~\cite{Cockayne2016}, using basis functions that satisfy the
underlying PDE, a probabilistic meshless method (PMM) is constructed.
In particular, if $\bm{\zeta}$ parameterizes positions of sources,
and $\phi(\v x,\bm{\zeta})=G(\v x,\bm{\zeta})$ in \eqref{eq:k12}
is chosen to be a fundamental solution / Green's function $G(\v x,\bm{\zeta})$
of the PDE, one may call the resulting scheme a \emph{probabilistic
method of fundamental solutions (pMFS)}. In~\cite{Cockayne2016}
sources are placed across the whole computational domain, and the
resulting kernel is called \emph{natural}. Here we will instead place
sources in the exterior to fulfill the homogeneous interior problem,
as in the classical MFS~\cite{Lackner1976,Golberg1995,Schaback2006}.
Technically, this is also achieved by setting $\Sigma_{\mathrm{p}}(\bm{\zeta},\bm{\zeta}^{\prime})=0$
for either $\bm{\zeta}$ or $\bm{\zeta}^{\prime}$ in the interior.
For discrete sources localized $\bm{\zeta}=\bm{\zeta}_{i}$ one obtains
again discrete basis functions $\phi_{i}(\v x)=G(\v x,\bm{\zeta}_{i})$
for~\eqref{eq:kbasis}.

More generally, according to theorem~2 of~\cite{van2001kriging},
for linear PDE operators $\hat{L}$ in \eqref{eq:Lu=00003Dq}~with
$q\neq0$ we require a Gaussian process of non-zero mean $m(\v x)$
with
\begin{align}
\hat{L}m(\v x) & =q(\v x),\label{eq:Lmisq}\\
\hat{L}k(\v x,\v x^{\prime}) & =0.\label{eq:kern}
\end{align}
Here $\hat{L}$ acts on the first argument of $k(\v x,\v x^{\prime})$.
Sources affect only the mean $m(\v x)$ of the Gaussian process, whereas
the kernel $k(\v x,\v x^{\prime})$ should be based on the homogeneous
equation. This hints to the technique of~\cite{OHagan1978} discussed
in~\cite{Rasmussen2006} chapter 2.7 to treat $m(\v x)$ via a linear
model added on top of a zero-mean process for the homogeneous equation.
In that case we consider is the superposition\\
\begin{align}
u(\v x) & =u_{h}(\v x)+u_{p}(\v x),\label{eq:hagan}\\
u_{h}(\v x) & \sim\mathcal{G}(0,k(\v x,\v x^{\prime})),\label{eq:uh}\\
u_{p}(\v x) & =\v h^{T}(\v x)\v b,\label{eq:up}\\
\v b & \sim\mathcal{N}(\v b_{0},B).
\end{align}
where $\v h^{T}(\v x)\v b$ is a linear model for $m(\v x)$ with
Gaussian prior mean $\v b_{0}$ and covariance $B$ for the model
coefficients. The homogeneous part~\eqref{eq:uh} corresponds to
a random process $u_{h}(\v x)$ where a source-free $k$ is constructed
according to~\eqref{eq:kern}. The inhomogeneous part~\eqref{eq:up}
may be given by any particular solution $u_{p}(\v x)$ for arbitrary
boundary conditions. Using the limit of a vague prior with $\v b_{0}=0$
and $|B^{-1}|\rightarrow0$, i.e. minimum information / infinite prior
covariance \cite{OHagan1978,Rasmussen2006}, posteriors for mean $\bar{u}$
and covariance matrix $\mathrm{cov}(u,u)$ based on given training
data $\v y=u(X)+\sigma_{\mathrm{n}}$ with measurement noise variance
$\sigma_{\mathrm{n}}^{2}$ are
\begin{align}
\bar{u}(X_{\star}) & =K_{\star}^{T}\,K_{y}^{-1}(\v y-H^{T}\bar{\v b})+H_{\star}^{T}\bar{\v b}=K_{\star}^{T}\,K_{y}^{-1}\v y+R^{T}\bar{\v b},\label{eq:expval}\\
\mathrm{cov}(u(X_{\star}),u(X_{\star})) & =K_{\star\star}-K_{\star}^{T}\,K_{y}^{-1}K_{\star}+R^{T}(HK_{y}^{-1}H^{T})^{-1}R.\label{eq:cova}
\end{align}
Here $X=(\v x_{1},\v x_{2},\dots\v x_{N})$ contains the training
points, $X_{\star}=(\v x_{\star1},\v x_{\star2},\dots,\v x_{\star N_{\star}})$
the evaluation or test points. Functions of $X$ and $X^{\star}$
are to be understood as vectors or matrices resulting from evaluation
at different positions, i.e. $\bar{u}(X_{\star})\equiv(\bar{u}(\v x_{\star1}),\bar{u}(\v x_{\star2}),\dots,\bar{u}(\v x_{\star N_{\star}}))$
is a tuple of predicted expectation values. The matrix $K\equiv k(X,X)$
is the kernel covariance of the training data with entries $K_{ij}\equiv k(\v x_{i},\v x_{j})$
and $\mathrm{cov}(u(X_{\star}),u(X_{\star}))_{ij}\equiv\mathrm{cov}(u(\v x_{\star i}),u(\v x_{\star j}))$
are entries of the predicted covariance matrix for $u$ evaluated
in the test points $\v x_{\star i}$. Furthermore $K_{y}\equiv k(X,X)+\sigma_{\mathrm{n}}^{\,2}I$,
$K_{\star}\equiv k(X,X_{\star})$, $K_{\star\star}\equiv k(X_{\star},X_{\star})$,
$R\equiv H_{\star}-HK_{y}^{-1}K_{\star\star}$, and entries of $H$
are $H_{ij}\equiv h_{i}(\v x_{j})$, $H_{\star ij}\equiv h_{i}(\v x_{\star j})$,
and $\bar{\v b}\equiv(HK_{y}^{-1}H^{T})^{-1}HK_{y}^{-1}\v y$.

\subsection{Linear modeling of sources}

A linear model for $m(\v x)$ fulfilling a PDE according to~\eqref{eq:kern}
follows directly from the source representation. Consider sources
to be modeled as a linear superposition over basis functions 
\begin{equation}
q(\v x)=\sum_{i}\ph_{i}(\v x)q_{i}
\end{equation}
with unknown source strength coefficients $\v q=(q_{i})$. To model
the mean instead of the source functions themselves, one uses an according
superposition
\begin{equation}
m(\v x)=\sum_{i}u_{p\,i}(\v x)q_{i}
\end{equation}
of particular solutions $u_{p\,i}(\v x)$ from inhomogeneous equations
\begin{equation}
\hat{L}u_{p\,i}(\v x)=\ph_{i}(\v x).\label{eq:upi}
\end{equation}
For the linear model~\eqref{eq:hagan} this means that $\v b=\v q$
and $h_{i}(\v x)=u_{p\,i}(\v x)$. Posterior mean of source strengths
and their uncertainty are
\begin{align}
\bar{\v q} & =(HK_{y}^{-1}H^{T})^{-1}HK_{y}^{-1}\v y,\label{eq:sourceest}\\
\mathrm{cov}(\v q,\v q) & =(HK_{y}^{-1}H^{T})^{-1}.\label{eq:sourceest2}
\end{align}
One can easily check that the predicted mean $\bar{u}(\v x_{\star})=\bar{u}_{h}(\v x_{\star})+\bar{u}_{p}(\v x_{\star})$
at a specific point $\v x_{\star}$ in~\eqref{eq:expval} fulfills
the linear differential equation~\eqref{eq:Lu=00003Dq}. In the homogeneous
part $\bar{u}_{h}(\v x_{\star})=k(\v x_{\star},X)\,K_{y}^{-1}(\v y-H^{T}\bar{\v q})$
sources are absent with $\hat{L}\bar{u}_{h}(\v x_{\star})=0$, with
$\hat{L}$ acting on $\v x^{\star}$ here. The particular solution
$\bar{u}_{p}(\v x_{\star})=\v h^{T}(\v x_{\star})\bar{\v q}=\sum_{i}u_{p\,i}(\v x_{\star})\bar{q}_{i}$
adds source contributions $q_{i}\varphi_{i}(\v x^{\star})$ due to~\eqref{eq:upi}.
For point monopole sources $\ph_{i}(\v x)=\delta(\v x-\v x_{q\,i})$
placed at at positions $\v x_{q\,i}$, the particular solution $u_{p,\,i}(\v x)$
equals the fundamental solution $G(\v x,\v x_{q\,i})$ evaluated for
the respective source. In the absence of sources the part described
in this subsection isn't modeled and (\ref{eq:expval}-\ref{eq:cova})
reduce to posteriors of a GP with prior mean $m(\v x)=0$ where matrix
$R$ vanishes.

\section{Application cases}

Here the general results described in the previous section are applied
to specific equations. Regression is performed based on values measured
at a set of sampling points $\v x_{i}$ and may also include optimization
of hyperparameters $\boldsymbol{\beta}$ appearing as auxiliary variables
inside the kernel $k(\v x,\v x^{\prime};\boldsymbol{\beta})$. The
optimization step is usually performed in a maximum a posteriori (MAP)
sense, choosing $\boldsymbol{\beta}_{\mathrm{MAP}}$ as fixed rather
than providing a joint probability distribution function including
$\boldsymbol{\beta}$ as random variables. We note that depending
on the setting this choice may lead to underestimation of uncertainties
in the reconstruction of $u$, in particular for sparse, low-quality
measurements.

\subsection{Laplace's equation in two dimensions}

First we explore construction of kernels in~\eqref{eq:uh} for a
purely homogeneous problem in a finite and infinite dimensional index
space, depending on the mode of separation. Consider Laplace's equation
\begin{equation}
\Delta u(\v x)=0.
\end{equation}
In contrast to the Helmholtz equation, Laplace's equation has no scale,
i.e. permits all length scales in the solution. In the 2D case using
polar coordinates the Laplacian becomes
\begin{equation}
\frac{1}{r}\frac{\partial}{\partial r}\left(r\frac{\partial u}{\partial r}\right)+\frac{1}{r^{2}}\frac{\partial^{2}u}{\partial\theta^{2}}=0.
\end{equation}
A well-known family of solutions for this problem based on the separation
of variables is
\begin{equation}
u=r^{\pm m}e^{\pm im\theta},
\end{equation}
leading to a family of solutions
\begin{equation}
r^{m}\cos(m\theta),\,r^{m}\sin(m\theta),\,r^{-m}\cos(m\theta),\,r^{-m}\sin(m\theta).
\end{equation}
Since our aim is to work in bounded regions we discard the solutions
with negative exponent that diverge at $r=0$. Choosing a diagonal
prior that weights sine and cosine terms equivalently \cite{Mendes2012}
and introducing a length scale $s$ as a free parameter we obtain
a kernel according to (\ref{eq:kbasis}) with
\begin{align}
k(\v x,\v x^{\prime};s) & =\sum_{m=0}^{\infty}\left(\frac{rr^{\prime}}{s^{2}}\right)^{m}\sigma_{m}^{\,2}\,(\cos(m\theta)\,\cos(m\theta^{\prime})+\sin(m\theta)\,\sin(m\theta^{\prime}))=\sum_{m=0}^{\infty}\left(\frac{rr^{\prime}}{s^{2}}\right)^{m}\sigma_{m}^{\,2}\,\cos\left(m(\theta-\theta^{\prime})\right).
\end{align}
A flat prior $\sigma_{m}^{\,2}=1$ for all polar harmonics and a characteristic
length scale $s$ as a hyperparameter, yields
\begin{align}
k(\v x,\v x^{\prime};s) & =\frac{1-\frac{rr^{\prime}}{s^{2}}\cos(\theta-\theta^{\prime})}{1-2\frac{rr^{\prime}}{s^{2}}\cos(\theta-\theta^{\prime})+\frac{\left(rr^{\prime}\right)^{2}}{s^{4}}}=\frac{1-\frac{\v x\cdot\v x^{\prime}}{s^{2}}}{1-2\frac{\v x\cdot\v x^{\prime}}{s^{2}}+\frac{|\v x|^{2}|\v x^{\prime}|^{2}}{s^{4}}}.\label{eq:laplace1}
\end{align}
This kernel is not stationary, but isotropic around a fixed coordinate
origin. Introducing a mirror point $\bar{\v x}^{\prime}$ with polar
angle $\bar{\theta}^{\prime}=\theta^{\prime}$ and radius $\bar{r}^{\prime}=s^{2}/r^{\prime}$
we notice that \eqref{eq:laplace1} can be written as
\begin{equation}
k(\v x,\v x^{\prime};s)=\frac{\v{\left|\bar{\v x}^{\prime}\right|}^{2}-\v x\cdot\v{\bar{\v x}^{\prime}}}{(\v x-\bar{\v x}^{\prime})^{2}},\label{eq:Kxxpr}
\end{equation}
making a dipole singularity apparent at $\v x=\bar{\v x}^{\prime}$.
In addition $k$ is normalized to $1$ at $\v x=0$. Choosing $s>R_{0}$
larger than the radius $R_{0}$ of a circle centered in the origin
and enclosing the computational domain, we have $\bar{r}^{\prime}>s^{2}/s=s>R_{0}$.
Thus all mirror points and the according singularities are moved outside
the domain.

\begin{figure}
\begin{centering}
\includegraphics[width=0.4\textwidth]{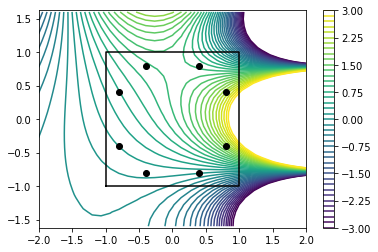}\includegraphics[width=0.4\textwidth]{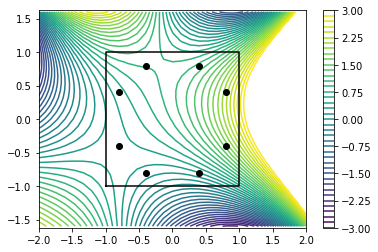}
\par\end{centering}
\begin{centering}
\includegraphics[width=0.4\textwidth]{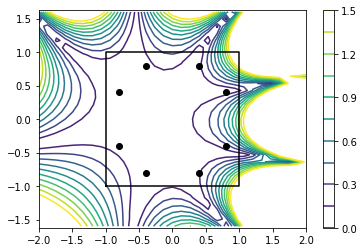}\includegraphics[width=0.4\textwidth]{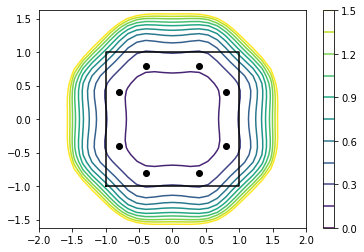}
\par\end{centering}
\caption{Analytical solution of Laplace equation (top left) and GP reconstruction
with source-free Mercer kernel~\eqref{eq:Kxxpr} (top right) with
absolute error (bottom left) and predicted 95\% confidence interval
(bottom right). Sources lie outside the black square region and measurement
positions are marked by black dots.\label{fig:Analytical-solution-of}}
\end{figure}

Choosing a slowly decaying $\sigma_{m}^{\,2}=1/m$, excluding $m=1$
and adding a constant term yields a logarithmic kernel instead \cite{Mendes2012}
with
\begin{equation}
k(\v x,\v x^{\prime};s)=1-\frac{1}{2}\ln\left(1-2\frac{\v x\cdot\v x^{\prime}}{s^{2}}+\frac{|\v x|^{2}|\v x^{\prime}|^{2}}{s^{4}}\right)=1-\ln\left(\frac{|\v x-\bar{\v x}^{\prime}|}{\v{\left|\bar{\v x}^{\prime}\right|}}\right).
\end{equation}
Instead of a dipole singularity that expression features a monopole
singularity at $\v x-\bar{\v x}^{\prime}$ that is avoided as mentioned
above.

Using instead Cartesian coordinates $x,y$ to separate the Laplacian
provides harmonic functions like
\begin{equation}
u=e^{\pm\kappa x}e^{\pm i\kappa y}.
\end{equation}
Here all solutions yield finite values at $x=0$, so we don't have
to exclude any of them \emph{a priori}. Introducing again a diagonal
covariance operator in~(\ref{eq:k12}) and taking the real part yields
\begin{align}
k(\v x,\v x^{\prime}) & =\int\varphi(\v x,\kappa)\sigma^{\,2}(\kappa)\varphi(\v x^{\prime},\kappa)\,\d\kappa=\mathrm{Re}\int_{-\infty}^{\infty}\sigma^{\,2}(\kappa)e^{\kappa(x\pm x^{\prime})}e^{i\kappa(y\pm y^{\prime})}\,\d\kappa.
\end{align}
Setting $\sigma^{\,2}(\kappa)\equiv e^{-2\kappa^{2}}$ and choosing
a characteristic length scale $s$ together with a possible rotation
angle $\theta_{0}$ of the coordinate frame yields the kernel
\begin{equation}
k(\v x,\v x^{\prime};s,\theta_{0})=\frac{1}{2}\mathrm{Re}\exp\left(\frac{\left((x+x^{\prime})\pm i(y-y^{\prime})\right){}^{2}e^{i2\theta_{0}})}{s^{2}}\right).
\end{equation}
Other sign combinations do not yield a positive definite kernel --
similar to the polar kernel~\eqref{eq:Kxxpr} before we couldn't
obtain an fully stationary expression that depends only on differences
between coordinates of $\v x$ and $\v x^{\prime}$.

For demonstration purposes we consider an analytical solution to a
boundary value problem of Laplace's equation on a square domain $\Omega$
with corners at $(x,y)=(\pm1,\pm1)$. The reference solution is
\begin{equation}
u_{\mathrm{ref}}(x,y)=\frac{1}{2}e^{y}\cos x+2x\cos(2y)
\end{equation}
and depicted in the upper left of Fig.~\ref{fig:Analytical-solution-of}
together with the extension outside the boundaries. This figure also
shows results from a GP fitted based on data with artificial noise
of $\sigma_{\mathrm{n}}=0.1$ measured at 8 points using kernel~\eqref{eq:Kxxpr}
with $s=2$. Inside $\Omega$ the solution is represented with errors
below $5\%$. This is also reflected in the error predicted by the
posterior variance of the GP that remains small in the region enclosed
by measurement points. The analogy in classical analysis is the theorem
that the solution of a homogeneous elliptic equation is fully determined
by boundary values.

\begin{figure}
\begin{centering}
\includegraphics[width=0.4\textwidth]{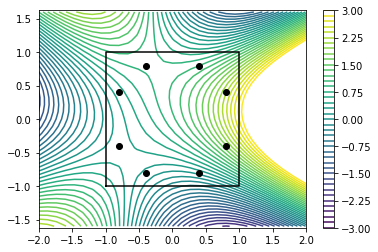}\includegraphics[width=0.4\textwidth]{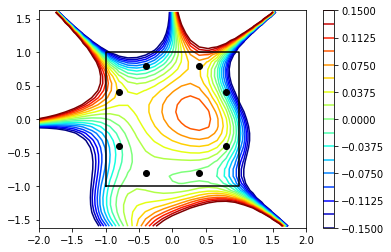}
\par\end{centering}
\begin{centering}
\includegraphics[width=0.4\textwidth]{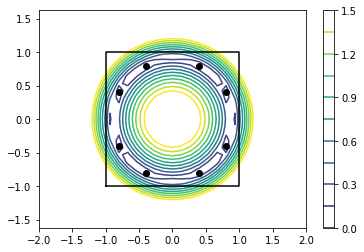}\includegraphics[width=0.4\textwidth]{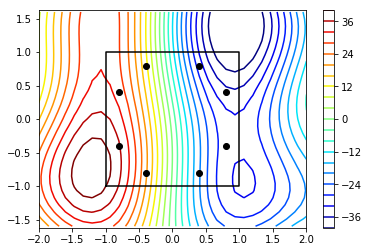}
\par\end{centering}
\caption{GP reconstruction of case in Fig.~\ref{fig:Analytical-solution-of}
with generic squared exponential kernel (top left) with predicted
95\% confidence interval (bottom left). Difference to reconstruction
with source-free kernel~\eqref{eq:Kxxpr} (top right) and source
density $\bar{q}=\Delta\bar{u}$ of prediction (bottom right).\label{fig:GP-reconstruction-of}}
\end{figure}

In comparison, a reconstruction using a generic squared exponential
kernel $k\propto\exp((\v x-\v x^{\prime})^{2}/(2s^{2}))$ yields a
result of similar approximation quality in Fig.~\ref{fig:GP-reconstruction-of}.
The posterior covariance of that reconstruction is however not able
to capture the vanishing error inside the enclosed domain due to given
boundary data. More severely, in contrast to the previous case, the
posterior mean $\bar{u}$ doesn't satisfy Laplace's equation $\Delta\bar{u}=0$
exactly. This leads to a violation of the classical result that (differences
of) solutions of Laplace's equation may not have extrema inside $\Omega$,
showing up in the difference to the reconstruction in Fig.~\ref{fig:GP-reconstruction-of}.
This kind of error is quantified by computation of the reconstructed
charge density $\bar{q}=\Delta\bar{u}$. This is fine if data from
Poisson's equation $\Delta u=q$ with distributed charges should be
fitted instead. However, to keep $\Delta u=0$ exact in $\Omega$,
one requires more specialized kernels such as~\eqref{eq:Kxxpr}.

\subsection{Helmholtz equation: source and wavenumber reconstruction}

To demonstrate the proposed method in full we now consider the Helmholtz
equation with sources
\begin{equation}
\Delta u(\v x)+k_{0}^{\,2}u(\v x)=q(\v x).\label{eq:helmi}
\end{equation}
Stationary kernels based on Bessel functions for the homogeneous equation
have been presented in~\cite{Albert2019a}. These functions provide
smoothing regularization on the order of the wavelength $\lambda_{0}=2\pi/k_{0}$
and have been demonstrated to produce excellent field reconstruction
from point measurements. Here we consider the two-dimensional case.
The method of source strength reconstruction is improved compared
to~\cite{Albert2019a}, as it constitutes a linear problem according
to~(\ref{eq:sourceest}-\ref{eq:sourceest2}). Non-linear optimization
is instead applied to wavenumber $k_{0}$ as a free hyperparameter
to be estimated during the GP regression.

\begin{figure}
\begin{centering}
\includegraphics[width=0.45\textwidth]{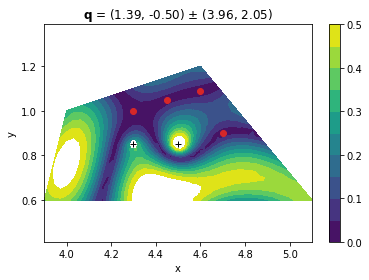}\includegraphics[width=0.45\textwidth]{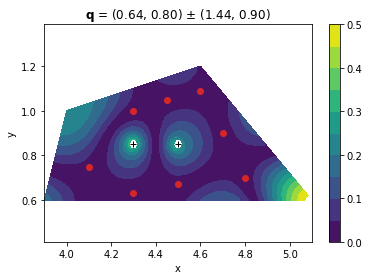}
\par\end{centering}
\begin{centering}
\includegraphics[width=0.45\textwidth]{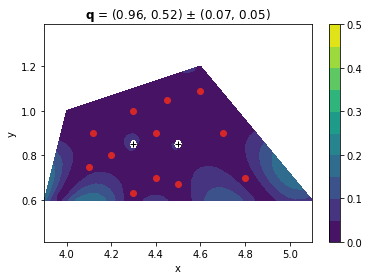}\includegraphics[width=0.45\textwidth]{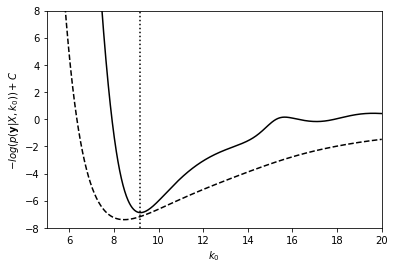}
\par\end{centering}
\caption{Reconstruction error for Helmholtz equation with different sensor
count (top, bottom left) and reconstructed source strengths $\protect\v q$
with $95\%$ confidence interval according to posterior (\ref{eq:sourceest}-\ref{eq:sourceest2}).
Negative log likelihood (bottom right) with optimum at $k_{0}^{\mathrm{ML}}=9.19$
for Bessel kernel~\cite{Albert2019a} (solid line), whereas the actual
value (dotted line) is $k_{0}=9.16$. The length scale of a squared
exponential kernel (dashed line) is less peaked. \label{fig:Source-strengths-}}
\end{figure}

The setup is the same as in~\cite{Albert2019a}: a 2D cavity with
various boundary conditions and two sound sources of strengths 0.5
and 1, respectively. Results for sound pressure fulfilling~\eqref{eq:helmi}
are normalized to have a maximum of $p/p_{0}=1$. Fig.~\ref{fig:Source-strengths-}
shows reconstruction error in field reconstruction depending on the
number of measurement positions. Here noise of $\sigma_{\mathrm{n}}=0.01$
has been added to the samples. The obtained negative log-likelihood
depending on $k_{0}$ permits an accurate reconstruction of this quantity
that has the physical meaning of a wavenumber. A generic squared exponential
kernel $k\propto\exp((\v x-\v x^{\prime})^{2}/(2(\pi/k_{0})^{2}))$
leads to results of similar quality and a slightly less peaked spatial
length scale hyperparameter without a direct physical interpretation.

\subsection{Heat equation}

Consider the homogeneous heat/diffusion equation
\begin{equation}
\frac{\partial u}{\partial t}-D\Delta u=0.
\end{equation}
for $(x,t)\in\mathbb{R}\times\mathbb{R}^{+}$. Integrating the fundamental
solution $G=1/\sqrt{4\pi(t-\tau)}\exp((x-\xi)^{2}/(4(t-\tau))$ from
$\xi=-\infty$ to $\infty$ at $\tau=0$, i.e. placing sources everywhere
at a single point in time, leads to the kernel
\begin{equation}
k_{\mathrm{n}}(x-x^{\prime},t+t^{\prime};D)=\frac{1}{\sqrt{4\pi D(t+t^{\prime})}}e^{-\frac{(x-x^{\prime})^{2}}{4D(t+t^{\prime})}}.\label{eq:kheatnat}
\end{equation}
In terms of $x$ this is a stationary squared exponential kernel and
the natural kernel over the domain $x\in\mathbb{R}$. The kernel broadens
with increasing $t$ and $t^{\prime}$. Non-stationarity in time can
also be considered natural to the heat equation, since its solutions
show a preferred time direction on each side of the singularity $t=0$.
The only difference of~\eqref{eq:kheatnat} to the singular heat
kernel is the positive sign between $t$ and $t^{\prime}$. If both
of them are positive, $k$ is guaranteed to takes finite values.

As for the Laplace equation it is also convenient to define a spatially
non-stationary kernel by cutting out a finite source-free domain.
Evaluating the integral over the fundamental solution in $\mathbb{R}\backslash(a,b)$
without our domain interval $(a,b)$ we obtain
\begin{equation}
k_{\mathrm{n}}(x,t,x^{\prime},t^{\prime})=k_{\mathrm{n}}(x-x^{\prime},t+t^{\prime};D)\left[1-\frac{g(x,t,x^{\prime},t^{\prime};D,b)-g(x,t,x^{\prime},t^{\prime};D,a)}{2}\right].
\end{equation}
where
\begin{equation}
g(x,t,x^{\prime},t^{\prime};D,s)\equiv\text{erf}\left(\frac{(s-x)/t+(s-x^{\prime})/t^{\prime}}{2\sqrt{D}\sqrt{1/t+1/t^{\prime}}}\right).
\end{equation}
Incorporating the prior knowledge that there are no domain sources
could potentially improve the reconstruction. Initial investigations
on the initial-boundary value problem of the heat equation based on
those kernels produce stable results showing natural regularization
within the limits of the strongly ill-posed setting. Reconstruction
of diffusivity $D$ has proven to be a difficult task and requires
further investigations.

\section*{Summary and Outlook}

A framework for application of Gaussian process regression to data
from an underlying partial differential has been presented. The method
is based on Mercer kernels constructed from fundamental solutions
and produces realizations that match the homogeneous problem exactly.
Contributions from sources are superimposed via an additional linear
model. Several examples for suitable kernels have been given for Laplace's
equation, Helmholtz equation and heat equation. Regression performance
has been shown to yield results of similar or higher quality to a
squared exponential kernel in the considered application cases. Advantages
of the specialized kernel approach are the possibility to represent
exact absence of sources as well as physical interpretability of hyperparameters.

In a next step reconstruction of vector fields via GPs could be formulated,
taking laws such as Maxwell's equations or Hamilton's equations of
motion into account. A starting point could be squared exponential
kernels for divergence- and curl-free vector fields~\cite{Narcowich1994}.
Such kernels have been used in \cite{Macedo2008} to perform statistical
reconstruction, and \cite{Cobb2018} apply them to GPs for source
identification in the Laplace/Poisson equation. In order to model
Hamiltonian dynamics in phase-space, vector-valued GPs could possibly
be extended to represent not only volume-preserving (divergence-free)
maps but retain full symplectic properties, thereby conserving all
integrals of motion such as energy or momentum.

\section*{Acknowledgments}

I would like to thank Dirk Nille, Roland Preuss and Udo von Toussaint
for insightful discussions. This study is a contribution to the \emph{Reduced
Complexity Models} grant number ZT-I-0010 funded by the Helmholtz
Association of German Research Centers.


\bibliographystyle{ieeetr}
\bibliography{Albert2019_Maxent}

\end{document}